# A Review of our Current Knowledge of Clouded Leopards (*Neofelis nebulosa*)

**Abstract**

Little is known about clouded leopards (*Neofelis nebulosa*), who have a vulnerable population that extends across southern Asia. We reviewed the literature and synthesized what is known about their ecology and behavior. Much of the published literature either note detections within and on the edges of their range, or are anecdotal observations, many of which are decades if not over a century old. Clouded leopards are a medium-sized felid, with distinctive cloud-shape markings, and notably long canines relative to skull size. Estimates for population densities range from 0.58 to 6.53 individuals per 100km2. Only 7 clouded leopards have been tracked via radio-collars, and home range estimates range from 33.6-39.7km$^2$ for females and 35.5-43.5km$^2$ for males. Most accounts describe clouded leopards as nocturnal, but radio telemetry studies showed that clouded leopards have arrhythmic activity patterns, with highest activity in the morning followed by evening crepuscular hours. There has never been a targeted study of clouded leopard diet, but observations show that they consume a variety of animals, including ungulates, primates, and rodents. We encourage future study of their population density and range to inform conservation efforts, and ecological studies in order to understand the species and its ecological niche.

Po-Jen Chiang[1]* and Maximilian L Allen[2]

[1]*Formosan Wild Sound Conservation Science Center, Taiwan*
[2]*University of Wisconsin, USA*

***Corresponding author:*** Po-Jen Chiang, Formosan Wild Sound Conservation Science Center Co. Ltd., No. 335, Yangmei, Taoyuan 32653, Taiwan, Email: neckrikulau@gmail.com

**Abbreviations:** SECR: Spatially Explicit Capture Recapture

## Population Status and Distribution

The global population of clouded leopards (*Neofelis nebulosa*) is considered vulnerable worldwide [1], and were considered to be less abundant globally the 2016 assessment by the IUCN than the previous assessment in 2007 [1]. Clouded leopards currently range from the southeastern Himalayas across southeastern Asia, extending through southern China and into peninsular Malaysia [2-6] (Figure 1). They have been extirpated from Taiwan [7], and possibly from Bangladesh [1]. Clouded leopards are protected in most countries across their range [1], but a large reason for population declines is unregulated hunting for pelts [1,8], as well as habitat fragmentation and regional declines in habitat quality [1].

Clouded leopards may be expanding their range recently in the southeastern Himalayas. Clouded leopards were historically found in Nepal [9], but were thought to have become extinct in the last century. They were rediscovered in the 1980's [5], and have now been documented as far west as Annapurna in central Nepal [10]. Similarly, clouded leopards were historically found in India and may have been common [11]. They are now much less common, and were documented for the first time in decades in the northeastern state of Mizoram [12], and were also detected during multiple surveys during recent decades in the northeastern state of Arunachal Pradesh [13,14], and Assam [15]. Future studies to document and confirm the global and local ranges of clouded leopards are important in order to set realistic and appropriate conservation goals. It is possible that the range is both expanding and contracting in different areas, making local conservation needs potentially different than those for the species as a whole. The expansions of the ranges of clouded leopards in the past few decades in Nepal and India [5,10,12-14], are encouraging. Although reintroductions are often difficult, because clouded leopards are naturally recolonizing areas it is possible that reintroduction efforts could also be successful. The causes of the original extirpation in the area need to be addressed and the long-term viability of the population considered before any reintroductions efforts begin.

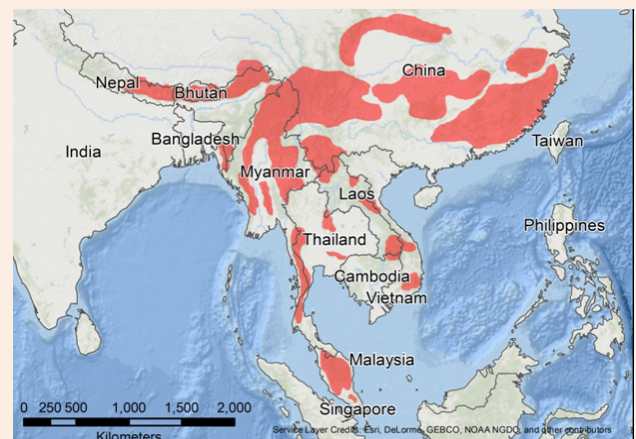

**Figure 1:** Distribution range (colored area) of clouded leopards, adapted from Grassman et al. [1].

## Population Densities and Abundance

The population density estimates of clouded leopards range from 0.58 to 6.53 individuals per 100 km$^2$ (Table 1). Advances in technology (i.e., motion-triggered cameras; [16]) and statistical methods (i.e. n-mixture models [15], and spatially-explicit capture recapture [SECR] models, [17]) make population densities easier to obtain in the past. As a result there has been a notable increase in the studies estimating clouded leopard population densities in the last decade. We strongly encourage this increase, with the goal of obtaining estimates of population densities across their entire range. This would allow researchers to understand baseline population densities in order to compare between populations, which is important for a species with vulnerable global status. These estimates can then be used to determine trends in population size and density over time. They can also be used to highlight areas of high conservation need, including where reintroductions may be feasible.

**Table 1:** The population density estimates available for clouded leopards from published literature, in number of individuals per 100 km$^2$, and including their location and statistical method of estimation.

| Study | Density | SE | Country | Statistical Method |
|---|---|---|---|---|
| Goswami & Ganesh | 0.58 | - | Assam, India | Relative Abundance Index |
| Borah et al. [6] | 5.5 | 1.8 | Assam, India | Jackknife Estimation in Capture |
| Borah et al. [6] | 6.53 | 1.88 | Assam, India | Pledger Models in Mark |
| Borah et al. [6] | 4.73 | 1.43 | Assam, India | Spatially-Explicit Capture Recapture |
| Mohamad et al. [88] | 3.46 | 1 | Peninsular Malaysia[1] | Spatially-Explicit Capture Recapture |
| Mohamad et al. [88] | 1.83 | 0.61 | Peninsular Malaysia[2] | Spatially-Explicit Capture Recapture |

Due to their cryptic nature of clouded leopards have been notoriously difficult to document in the past, but the development of commercial camera traps has increased the ability for researchers to document clouded leopards (Table 2). Many of the past population estimates from camera data have large error values, however, and an area of research need is to determine methods of increasing accuracy. Methods of increasing detection rates, including the possible use of lures (e.g., [18]) and increasing the accuracy of the target locations should be considered. Another solution is to increase the density of cameras, along with possibly reducing the overall grid size, in order to increase the capture effectiveness. This may allow for greater capture and recapture of individuals, which would decrease the error values in population estimates. In SECR models, however, the maximum size between cameras should be the radius of a home range [17]. The 2km distance used by Borah is well below the radius of known home ranges of clouded leopards (see below), although home range size could vary across their range. In conclusion, although camera density is important to consider, other methods to increase detection rates may be more important for decreasing the error estimates of population densities.

## Physical and Genetic Characteristics

Although belonging to *Pantherinae*, clouded leopards are a medium-sized felid, weighing between 11-23kg [4,19-21]. They have distinctive large dark, cloud-shape markings, a tail typically as long as its head-body length (up to 80-90cm: [19-22]. Clouded leopards are notable for their relatively the long canine relative to skull size (3.8-4.5cm: [3]), the longest of living felids and reminiscent of saber-toothed tigers [23-24]. Although the skull of clouded leopards does not reach pantherine size, it has attained pantherine cranial proportions (especially large teeth) [25]. Clouded leopards also have behavioral characteristics that fall between those of large and small felids [3,26]. For example, similar to small felid species, they purr and cannot roar. Their method of eating food, grooming, and its body postures, however, are closer to those of the larger species of felids [26,27].

**Table 2**: The camera trapping effort estimates available for clouded leopards from published literature with >3 photos, with effort estimates are in mean number of days for a capture.

| Study | Effort Days | Country |
|---|---|---|
| Datta et al. [13] | 768 | Arunachal Pradesh, India |
| Goswami & Ganesh | 272.3 | Assam, India |
| Mohamad et al. [88] | 128.8 | Peninsular Malaysia[1] |
| Mohamad et al. [88] | 149.2 | Peninsular Malaysia[2] |
| Moo et al. [75] | 528.4 | Myanmar |
| Ngoprasert et al. [74] | 567.7 | Thailand |
| Tan et al. [75] | 173.2 | Peninsular Malaysia |
| Tempa et al. [94] | 113.1 | Bhutan |

Originally, clouded leopards were classified as a single member of the genus *Neofelis* [28] that consisted of 4 subspecies [4]. However, the latest genetic and morphometric research suggested that the former subspecies *N.n. diardi* in Sumatra and Borneo should be classified as a distinct species (i.e. *N. diardi*) based on distinct haplotypes [29], and pelage (smaller cloud markings, more cloud spots, and greyer fur) [30]. The Formosan clouded leopard (*Neofelis nebulosa brachyurus*) in Taiwan was first introduced to scientists by Swinhoe in 1862 [2] and was described as a distinct species, *Leopardus brachyurus*, based on a shorter tail length [2]. But [31], revised it to an insular race of the continental clouded leopards after acquiring more specimens. [32,19] maintained that tail length is not a consistent criterion. [33,34]) even suggested that it is unnecessary to classify the Formosan clouded leopard as a distinct subspecies. Using samples from the National Taiwan

Museum (7 specimen, but DNA was successfully extracted from only 1 sample), the latest genetic analysis showed that Taiwan clouded leopards diverged from the other mainland subspecies in haplotypes, but not to the level of a distinct species [29].

Heterozygosity within clouded leopards has been examined in a population of 20 captive animals from USA zoos using allozymes only. The percent average heterozygosity (H) of clouded leopards was 2.3 [35], which are similar to 2.3 for free ranging lions (*Panthera leo*) in Kruger National Park [36,37]. However, clouded leopards had the fewest number of allozymes polymorphisms compared to 9 other felid species, with only the cheetah (*Acinonyx jubatus*) (a known bottleneck species) showing less heterozygosity [35,36]. Further genetic work is needed to understand heterozygosity in given populations.

### Home Range and Movement Patterns

Over the last two decades researchers placed the first radio tracking collars on free-ranging clouded leopards [38-39]. Tracked 2 adult clouded leopards in Thailand and reported that they occupied similarly sized home ranges (39.5km$^2$ for 1 female and 42.2km$^2$ for 1 male, 95% fixed kernel) [39]. Results in Thailand also showed no obvious differences of home range size (95% fixed kernel) between 2 adult males (35.5 and 43.5km$^2$) and 2 adult females (33.6 and 39.7km$^2$) in another area in Thailand. In addition, during camera trapping estimates from India individual clouded leopards were detected at cameras placed 2.0 to 13.6km apart. It is difficult to draw conclusions from such small sample sizes, but we note 2 observations of interest. First, the home range size of male and female clouded leopards appear to be similar in size, although male solitary felids tend to have notably larger home range sizes than females [40-44]. Second, although there is a positive correlation between home range size and body size [45-47], the home range size reported for clouded leopards in Thailand [38-39] are twice as large as male leopard (*Panthera pardus*) home ranges (18km$^2$) reported elsewhere in Thailand [48]. We are not sure if the reported home ranges are accurate for clouded leopards across their range, or if they are affected by local factors. A large variation in home range size has been observed in many solitary felid species, with a 10-fold variation in HRS common for species including bobcats (*Lynx rufus*) [40], jaguars (*Panthera onca*) [41,49,50], leopards [51,42,44] and pumas (*Puma concolor*) [43-52]. Based on this potential variation, we do not know where the current estimates of clouded leopard home range size fall on the spectrum. Many of these studies have concluded that prey distribution and abundance, at least in part, is associated with home range size and population dynamics [40,44,50,53,54].

Austin in 2002 tracked 2 radio-collared clouded leopards and reported that the female had a mean daily movement distance of 977m; the male had a mean daily movement distance of 1,168m, while Grassman LI [39] reported an average 1,932m for 4 radio-tracked clouded leopards (range 122-7,724m). These are based on straight-line measurements and the distance moved could be higher when animals meandered between sampling locations. No dispersal data about clouded leopards are available. One subadult male clouded leopard was caught by local villagers in Nepal, and was then radio-collared and translocated 100 km east of the original capture site [5]. During the first 8 days of tracking it occupied an area less than 1 km$^2$, and then moved west toward where it was originally captured before the collar fell off after 10 days [5]. Considering the large range of clouded leopards there are likely many factors that affect home range size, movement patterns, and dispersal. We are a proponent of non-invasive studies whenever possible, however, placing more radio-collars on clouded leopards across their range would increase our understanding of home range size and movement patterns. This would inform conservation efforts and likely also contribute to our understanding of other aspects of their ecology.

### Reproductive and Communication Behavior

Reproductive and communication behavior is unknown in the wild for clouded leopards. A study of Sunda clouded leopard showed that they were similar to other felids (e.g., [55,56]) in using scent marking for intraspecific communication including territoriality, male-male competition, and mate selection [57]. Sunda clouded leopards used 10 communication behaviors, including scraping, claw marking, urine spraying and cheek rubbing [57]. Male Sunda clouded leopards ranged across larger areas, and overlapped at scent marking areas, while females did not [57].

Clouded leopards are likely to exhibit reproductive and communication behaviors similar to other felids, especially Sunda clouded leopards and other species of the *Panthera* lineage Sensu [57]. We encourage the use of motion-triggered cameras, which are a good method for documenting these behaviors and can be used across the range of clouded leopards. In captivity, clouded leopards begin breeding at 2 years old, and gestation lasts about 90 days [58]. Clouded leopards can exhibit seasonality in breeding depending on length of day and light availability [58]. Estrus lasts 3 to 6 days, and clouded leopards can exhibit spontaneous ovulation [58]. Female clouded leopards exhibit increases in scent marking prior to estrus [58], which is similar to other solitary felids [59]. Although studies in captivity inform our biological understanding, captivity also changes behavior [60,61], and understanding reproduction and communication behaviors in natural systems is an important area of research needed in the future.

### Arboreal Behavior

Clouded leopards have arboreal talents that rival those of the margay (*Leopardus wiedi*) [4]. Their relatively short, but powerful legs, large feet, and long tail are adaptations for arboreal life, giving clouded leopards a low center of gravity and a good grip on tree branches [21,62-64]. In captivity, clouded leopards have been observed to climb about on horizontal branches with its back to the ground, and hang upside down from branches by its hind feet [65]. Such behavior has been related to the hunting method of clouded leopards, in which they hang over tree branches and jump down upon passing prey [21]. They have also been observed running down tree trunks headfirst in captivity [65], and was once observed in the wild hunting among a troop of pigtail macaques (*Macaca nemestrina*) [66].

Because of their arboreal talents, most literature describes clouded leopards as mainly arboreal based on local surveys and captive observations [20,21,67-71]. Findings from studies using radio telemetry, however, suggested that clouded leopards may travel on the ground more often than in the trees [5, 39,72]. Contended that it could be difficult for clouded leopards to travel long distances through the trees. However, comparing the ratio of the sighting records in trees, clouded leopards are likely not primarily arboreal and instead uses trees as resting and hunting sites [3,66,73] but variations may occur in different habitats or regions.

### Habitat Use

Early literature indicated that clouded leopards occurred in dense primary forests [19,20,67,68]. Recent information, however, shows that clouded leopards are versatile and could occur in many different habitats. These habitats include: grassland [5], secondary or selectively logged forests [71], mature evergreen rain forests [71,74]. These accounts however are opportunistic, and based on local interviews and records from hunting, tracks, and direct observations. Radio telemetry studies in Thailand showed variations in forest use comparing closed primary forest and more open secondary forest-grassland habitat [38,39]. Three of the six clouded leopards tracked used vegetation types proportionally and two preferred closed primary forest. One occurred more in the open forest-grassland, which led [39] to suggest that this particular clouded leopard used edges as hunting sites. Their results provided support for the generally held belief that clouded leopards occur in primary evergreen forest [4]. Camera trapping in peninsular Malaysia has shown a preference in clouded leopards for forested habitats, and detections were more frequent with increasing distance from water bodies [75].

Much of the literatures suggests that clouded leopards occur most often in lowlands [19,66,68,70,17]. Other studies show that clouded leopards could occur as high as 2,585m in northeastern India [71], and clouded leopard detection at cameras was associated with higher elevations than lower elevation in Thailand [74]. In peninsular Malaysia clouded leopard detections at cameras increased at higher elevations [75]. However, occurrences of clouded leopards at these higher altitudes were extremely rare in the literature and were mostly indirect records based on interviews except a sighting by biologists at altitude 2,157m in northeastern India [12], and being found at nearly 3,000m in Nepal [10]. Future studies of habitat based on explicit data are needed to understand the habitat clouded leopards are selecting for, and what drives this selection [74]. Found that clouded leopard occurrence was linked to habitat used by red muntjac (*Muntiacus muntjak*) and Eurasian wild pigs (*Sus scrofa*), but further studies are needed to understand if habitat use is linked to habitat preferred by prey, or other factors such competition refuges from larger carnivores (e.g., [76,15]), and avoidance of people. Habitat use and elevation preferences are needed for setting realistic and effective conservation goals.

### Activity Patterns

Most accounts describe clouded leopards as nocturnal due to rare observation [2,19,21,68-70,77]. Since clouded leopards have sometimes been seen traveling or hunting during daytime [66], they may not be as strictly nocturnal as previously assumed. Radio telemetry studies in Thailand showed that clouded leopards have arrhythmic activity patterns [38,39], with highest activity in the morning (801-1200) followed by evening crepuscular hours (1801-2000) [39]. Camera trapping in Thailand showed that 73% of observations were nocturnal [78]. A future area of research is to determine the activity patterns of clouded leopards, if they vary in different areas, and if so, what causes the variation.

Two variables that may affect the activity patterns of clouded leopards are the activity patterns of prey and dominant carnivores [79]. Proposed that predators track the activity periods of their prey, the activity patterns of felids, including ocelots (*Leopardus pardalis*), jaguars, and pumas, are related to those of their prey [80]. Felids are also known to shift their temporal activity to avoid times when dominant carnivores are active, and in Thailand times of activity for clouded leopards most overlapped leopard cats (*Prionailurus bengalensis*) and tigers (*Panthera tigris*), with least overlap of leopards and marbled cats (*Pardofelis marmorata*) [78]. The use of radio tracking collars will be able to document activity patterns of clouded leopards, while the use of motion-triggered cameras could be used to determine activity patterns of clouded leopards as well as their prey and any dominant carnivores in the area.

### Food Habits

Like many other felids, clouded leopards consume a variety of animals, including ungulates, primates, rodents, domestic animals, and sometimes fish and snakes (Table 3). Our current understanding of clouded leopard diet is primarily from isolated observations, and is sure to be incomplete and not entirely accurate. For example, domestic animals are frequently reported, but this is likely because people are most likely to observe clouded leopards feeding on domestic animals. In contrast to Sunda clouded leopards, no birds have yet been noted in the diet of clouded leopards, but this is likely due to the very small sample of observations of clouded leopard prey to date. Also, although [39] reported small mammals such as the Indochinese ground squirrel (*Menetes berdmorei*) and *Muridae* species in the diet, the stocky build, large canines and the large post canine space make clouded leopards capable of killing relatively large prey [19,21,81]. It has historically been reported that clouded leopards will return to unfinished kills [21,82,83] recently confirmed this by discovering a dead domestic goat cached on a tree branch 4m above the ground and saw a clouded leopard return to the kill the next day. Returning to feed on prey for multiple days until it is consumed is common in solitary felids (e.g., [42,52], especially when killing large prey that take multiple days to consume. Considering how important feeding ecology is to carnivores, remarkably little is known about the diet and feeding patterns of clouded leopards. We encourage future studies on the feeding ecology of clouded leopards, including diet composition, duration of feeding bouts, and how diet affects the energetic of clouded leopards. These studies could easily tie in with studies of activity patterns and how they are affected by prey species, how prey availability affects home range size, and how prey density affects the population densities of clouded leopards.

**Table 3:** The known prey items of clouded leopards. Studies are listed in chronological order, with their location, type of study, and whether the known frequent prey types were observed during the study. In the footnotes we list the known food items in as much detail as possible from the given study. Prey items. 1. Goats and Pigs, 2. Deer, 3. Cattle, 4. Porcupine, 5. Pig-Tailed Macaque (*Macaca Nemestrina*), 6. Goats, 7. Hog Deer (*Axis porcine*), 8. Malayan Pangolin (*Manis javanica*), 9. Slow Loris (*Nycticebus coucang*), 10. Bush tailed porcupine (*Atherurus macrourus*), 11. Indochinese ground squirrel (*Menetes berdmorei*).

| Source | Location | Type | Ungulates | Primates | Birds | Domestic | Rodent |
|---|---|---|---|---|---|---|---|
| Tickell [67] | India | Observation | No | No | No | Yes | No |
| Swinhoe [2] | Formosa | Observation | Yes |  | No | No | No |
| Brownlow | Tavoy | Observation | No | No | No | Yes | No |
| Editors | India | Observation | No | No | No | No | Yes |
| Davies [66] | Thailand | Observation | No | Yes | No | No | No |
| Hazarika | Assam | Observation | No | No | No | Yes | No |
| Grassman et al. [39] | Thailand | Observation | Yes | No | No | No | No |
| Grassman et al. [39] | Thailand | Scat/Trap | No | Yes | No | No | Yes |

## Recommendations for Future Studies

The ecology and behavior of clouded leopards is relatively unknown when compared to other *Panthera* species. However, their populations are vulnerable world-wide [1], and are split into distinct sub-populations, many of which are declining [1]. Considering their vulnerable status and their standing as a *Panthera* species, which are often flagship species for conservation [84], they are a species that needs more study. Many of the published studies about clouded leopards either a) note detections within and on the edges of their range, or b) are anecdotal studies recording observations, many of which are decades if not over a century old. Only 7 clouded leopards have ever had radio-collars placed on them, and these individuals are restricted to a small portion of the distribution of clouded leopards [85-92].

## Conclusion

Our areas of emphasis for future work include: 1) Population density and range. We consider an increased number of studies on clouded leopard population densities to be essential in future conservation of the species. The use of camera traps and new statistical methods appear to be good methods for estimating densities, and can be applied across their range. 2) Home range and movement. In order to understand clouded leopards we need to understand basic ecology, such as how much space they use, and understanding movement patterns will also be important for creation of corridors for conservation. 3) Ecological studies. There is much room for ecological studies, with recent studies have allowed us to remove folklore such as clouded leopards being primarily arboreal. However, little is still known about prey and hunting tactics, energetic, scent marking and communication, or interactions with other carnivores. Understanding these areas of clouded leopard natural history and ecology will help in creating effective conservation goals, and understanding the species and its ecological niche.

## Acknowledgement

The authors declare that there are no Acknowledgements regarding the publication of this manuscript.

## Conflict of Interest

## References

1. Mariel Grassmann, Elke Vlemincx, Andreas von Leupoldt, Justin M Mittelstädt, Omer Van den Bergh (2016) Respiratory Changes in Response to Cognitive Load: A Systematic Review. Neural Plasticity 8146809.

2. Swinhoe R (1862) On the mammals of the island of Formosa. Proceedings of the Zoological Society of London 23: 347-365.

3. Guggisberg CAW (1975) Wild cats of the world. (1st edn), Taplinger Pub. Co, New York, USA.

4. Nowell K, Jackson P (1996) Wild cats: status survey and conservation action plan. IUCN, Gland, Switzerland.

5. Dinerstein E, Mehta JN (1989) The clouded leopard in Nepal. Oryx 23: 199-201.

6. Borah J, Sharma T, Das D, Rabha N, Kakati N, et al. (2013) Abundance and density estimates for common leopard *Panthera pardus* and clouded leopard Neofelis nebulosa in Manas National Park, Assam, India. Oryx 48: 149-155.

7. Chiang CE, Wang TD, Ueng KC, Lin TH, Yeh HI, et al. (2015) Guidelines of the Taiwan Society of Cardiology and the Taiwan Hypertension Society for the management of hypertension. J Chin Med Assoc 78(1): 1-47.

8. D Cruze N, Macdonald DW (2015) Clouded in mystery: the global trade in clouded leopards. Biodiversity and Conservation 24(14): 3505-3526.

9. Hodgson BH (1853) *Felis macrosceloides.* Proceedings of the Zoological Society of London, BHL, London.

10. Ghimirey Y, Acharya R, Adhikary B, Werhahn G, Appel A (2013) Clouded leopard camera-trapped in the Annapurna Conservation Area, Nepal. Cat News 58: 25.

11. Jerdon TC (1874) The mammals of India: a natural history of all the animals known to inhabit continental India. London, J. Wheldon, London.

12. Ghose D (2002) First sighting of clouded leopard *Neofelis nebulosa* from the Blue Mountain National Park, Mizoram, India. Current Science 83(1): 20-21.

13. Datta A, Anand MO, Naniwadekar R (2008) Empty forests: Large carnivore and prey abundance in Namdapha National Park, northeast India. Biological Conservation 141: 1429-1435.

14. Velho N, Srinivasan U, Singh P, Laurance WF (2016) Large mammal use of protected and community-managed lands in a biodiversity hotspot. Animal Conservation 19: 199-208.

15. Royle JA (2004) N-mixture models for estimating population size from spatially replicated counts. Biometrics 60(1): 108-115.

16. Burton AC, Neilson E, Moreira D, Ladle A, Steenweg R, et al. (2015) Wildlife camera trapping: A review and recommendations for linking surveys to ecological processes. Journal of Applied Ecology 52: 675-685.

17. J Royle, Richard B Chandler, Rahel Sollmann, Beth Gardner (2013) Spatial capture-recapture. Academic Press, USA.

18. Tanner D, Zimmerman P (2012) Optimal attractants to increase visits by clouded leopards to remote-camera sets. Wildlife Society Bulletin 36(3): 594-599.

19. Pocock RI (1939) The fauna of British India. (2nd edn), Taylor and Francis, London.

20. Prater SH (1965) The book of Indian animals, (2nd edn), Bombay Natural History Society: Prince of Wales Museum of Western India, India.

21. Lekagul B, MacNeely JA, Marshall JT, Askins R (1977) Mammals of Thailand. Sahakarnbhat, Thailand.

22. Metha JN, Dhewaju RG (1990) A note on the record of clouded leopards in Nepal. AGRIS 17: 21-22.

23. Sterndale RA (1884) Natural history of the mammalia of India and Ceylon. Himalayan Books, India.

24. Christiansen P (2006) Sabertooth characters in the clouded leopard (*Neofelis nebulosa* Griffiths 1821). J Morphol 267(10): 1186-1198.

25. Werdelin L (1983) Morphological patterns in the skulls of cats. Biological Journal of the Linnean Society 19(4): 375-391.

26. Gao YT (1987) Fauna Sinica, Mammalia, Vol. 8: Carnivora. Science Press, China.

27. Mellen JD (1991) Little-known cats, Great cats: majestic creatures of the wild. In: Seidensticker J & Lumpkin S, et al. (Eds.), Rodale Press, USA, pp. 170-179.

28. Ewer RF (1973) The carnivores. Comstock Pub Associates. USA.

29. Buckley Beason VA, Johnson WE, Nash WG, Stanyon R, Menninger JC, et al. (2006) Molecular evidence for species-level distinctions in clouded leopards. Curr Biol 16(23): 2371-2376.

30. Kitchener AC, Beaumont MA, Richardson D (2006) Geographical variation in the clouded leopard, *Neofelis nebulosa*, reveals two species. Curr Biol 16(23): 2377-2383.

31. Swinhoe R (1870) Catalogue of the mammals of China (south of the river Yangtsze) and the island of Formosa. Proceedings of the Zoological Society of London 32: 615-653.

32. Horikawa Y (1930) Survey of Formosan mammals (IV). Transactions of the Natural History Society of Formosa 20: 277-284.

33. Kuroda N (1938) Checklist to the mammals of Japan, Japan.

34. Kuroda N (1940) A monograph of the Japanese mammals. The Sanseido Co. Ltd., Tokyo, Japan.

35. Wang Y, Chu S, Wildt D, Seal U S (1995) Clouded leopard-Taiwan (*Neofelis nebulosa brachyurus*) Population and Habitat Viability Assessment. IUCN/SSC Conservation Breeding Specialist Group.

36. Newman A, Bush M, Wildt DE, Vandam D, Frankenhuis MT, et al. (1985) Biochemical genetic variation in 8 endangered or threatened felid species. Journal of Mammalogy 66: 256-267.

37. Miththapala S, Seidensticker J, Phillips LG, Goodrowe KL, Fernando SBU, et al. (1991) Genetic variation in Sri-Lankan leopards. Zoo Biology 10: 139-146.

38. Austin SC (2003) Ecology of sympatric carnivores in Khao Yai National Park, Thailand. The university of Hong Kong Libraries.

39. Grassman LI, Tewes ME, Silvy NJ, Kreetiyutanont K (2005) Ecology of three sympatric felids in a mixed evergreen forest in north-central Thailand. Journal of Mammalogy 86: 29-38.

40. Bailey TN (1974) Social organization in a bobcat population. JSTOR 38(3): 435-446.

41. Schaller GB, Crawshaw PG (1980) Movement patterns of jaguar. JSTOR 12(3): 161-168.

42. Bailey TN (1993) The African leopard: ecology and behavior of a solitary felid. The university of Chicago press journals 69(4).

43. Brian R, Spreadbury RR, Kevin Musil, Jim Musil, Chris Kaisner, et al. (1996) Cougar population characteristics in southeastern British Columbia. JSTOR 60(4): 962-969.

44. Mizutani F, Jewell PA (1998) Home-range and movements of leopards (*Panthera pardus*) on a livestock ranch in Kenya. AGRIS 244: 269-286.

45. Harestad AS, Bunnell FL (1979) Home range and body weight - re-evaluation. Ecology 60(2): 389-402.

46. Gittleman JL, Harvey PH (1982) Carnivore home-range size, metabolic needs and ecology. JSTOR 10(1): 57-63.

47. Mace GM, Harvey PH, Clutton Brock TH (1983) Vertebrate home-range size and energetic requirements. In: Swingland IR & Greenwood PJ (Eds.), The ecology of animal movement. Clarendon Press, UK, p. 32-53.

48. Grassman LI (1999) Ecology and behavior of the Indochinese leopard in Kaeng Krachan National Park, Thailand. Natural History Bulletin of the Siam Society 47: 77-93.

49. Rabinowitz AR, Nottingham BG (1986) Ecology and behavior of the jaguar (*Panthera onca*) in Belize, Central-America. Journal of Zoology 210: 149-159.

50. Crawshaw PG, Quigley HB (1991) Jaguar spacing, activity and habitat use in a seasonally flooded environment in Brazil. Journal of Zoology 223: 357-370.

51. Bertram BCR (1982) Leopard ecology as studies by radio tracking. SSC 49: 341-352.

52. Elbroch LM, Lendrum P, Allen ML, Wittmer HU (2015) Nowhere to hide: pumas, black bears, and competition refuges. Behavioral Ecology 26(1): 247-254.

53. Seidensticker J (1976) On the ecological separation between tigers and leopards. JSTOR 8(4): 225-234.

54. Emmons LH (1988) A field study of ocelots (Felis pardalis) in Peru. Revue D Ecologie 43: 133-157.

55. Smith JLD, McDougal C, Miquelle D (1989) Scent marking in free-ranging tigers, Panthera tigris. Animal Behaviour 37: 1-10.

56. Allen ML, Hocevar L, Krofel M (2017) Where to leave a message? The selection and adaptive significance of scent-marking sites for Eurasian lynx. Behavioral Ecology and Sociobiology 71: 136.

57. Maximilian L Allen, Heiko U Wittmer, Endro Setiawan, Sarah Jaffe, Andrew J Marshall, et al. (2016) Scent marking in Sunda clouded leopards (*Neofelis diardi*): novel observations close a key gap in understanding felid communication behaviours. Sci Rep 6: 35433.

58. Brown JL (2011) Female reproductive cycles of wild female felids. Animal Reproduction Science 124(3): 155-162.

59. Allen ML, Wittmer HU, Houghtaling P, Smith J, Elbroch LM, Wilmers CC (2015) The role of scent marking in mate selection by female pumas (*Puma concolor*). PLoS One 10: e0139087.

60. Mallapur A, Chellam R (2002) Environmental influences on stereotypy and the activity budget of Indian leopards (*Panthera pardus*) in four zoos in southern India. Zoo Biology 21: 585-595.

61. Quirke T, O Riordan RM, Zuur A (2012) Factors influencing the prevalence of stereotypical behaviour in captive cheetahs (*Acinonyx jubatus*). AGRIS 142: 189-197.

62. Gonyea WJ (1976) Adaptive differences in the body proportions of large felids. Acta Anat (Basel) 96(1): 81-96.

63. Gonyea WJ (1978) Functional Implications of felid forelimb anatomy. Acta Anatomica 102: 111-121.

64. Taylor ME (1989) Locomotor adaptions by carnivores. In: JLE Gittleman (Ed.), Carnivore behavior, ecology, and evolution. Comstock, Greece, pp. 382-409.

65. Hemmer H (1968) Untreated by the Pantherkatzen (Pantherinae) II: Studien zur ethologie des Nebelparders Neofelis nebulosa (Griffith 1821) and Irbis Uncia uncia (Schreber 1775). [Studies of the phylogenetic history of the Pantherina II: Research into the ecology of the clouded leopard and snow leopard.]. Veröffentlichungen der Zoologischen Staatssammlung München 12: 155-247.

66. Davies RG (1990) Sighting of a clouded leopard (*Neofelis nebulosa*) in a troop of pigtail macaques (*Macaca nemestrina*) in Khao Yai National Park, Thailand. Siam Society Natural History Bulletin 38: 95-96.

67. Tickell SR (1843) Notes on a curious species of tiger or jaguar, killed near the Snowy Range, north of Darjeeling. Journal of the Asiatic Society of Bengal 12: 814-816.

68. Renshaw G (1905) More natural history essays. Sherratt and Hughes, London.

69. Humphrey SR, Bain JR (1990) Endangered animals of Thailand. Sandhill Crane Press, USA.

70. Choudhury A (1993) The clouded leopard in Assam. Oryx 27(1): 51-53.

71. Choudhury A (1997) The clouded leopard in Manipur and Nagaland. Journal of the Bombay Natural History Society 94: 389-391.

72. Austin SC, Tewes ME (1999) Ecology of the clouded leopard in Khao Yai National Park, Thailand. Cat News 31: 17-18.

73. Lloyd E, Kreeltiyutanont K, Prabnasuk J, Grassman LI, Borries C (2006) Observation of Phayre's leaf monkeys mobbing a clouded leopard at Phu Khieo Wildlife Sanctuary (Thailand). Mammalia 70(12): 158-159.

74. Ngoprasert D, Lynam AJ, Sukmasuang R, Tantipisanuh N, Chutipong W, et al. (2012) Occurrence of three felids across a network of protected areas in Thailand: prey, intraguild, and habitat associations. Biotropica: 1-8.

75. Tan CKW, Rocha DG, Clements GR, Brenes Mora E, J Wadey J (2017) Habitat use and predicted range for the mainland clouded leopard Neofelis nebulosa in Peninsular Malaysia. Biological Conservation 206: 65-74.

76. Durant SM (1998) Competition refuges and coexistence: an example from Serengeti carnivores. JSTOR 67(3): 370-386.

77. Kanchanasakha B, Simcharoen S, Than UT (1998) Carnivores of mainland South-East Asia. (1st Edn) WWF, Bangkok.

78. Lynam AJ, Jenks KE, Tantipisanuh N, Chutipong W, Ngoprasert D, et al. (2013) Terrestrial activity patterns of wild cats from camera-trapping. Raffles Bulleting of Zoology 61: 407-415.

79. Curio E (1976) The ethology of predation. New York, Berlin.

80. Emmons LH (1987) Comparative feeding ecology of felids in a neotropical rainforest. JSTOR 20(4): 271-283.

81. Therrien F (2005) Feeding behaviour and bite force of sabretoothed predators. Zoological Journal of the Linnean Society 145: 393-426.

82. Kano T (1930) The distribution and habit of mammals of Formosa. Zoological Magazine 42: 165-173.

83. Hazarika AA (1996) Goat predation by clouded leopard (*Neofelis nebulosa*) in Kakoi Reserve Forest of Assam. Biodiversity Heritage Library 93: 584-585.

84. Belbachir F, Pettorelli N, Wacher T, Belbachir Bazi A, Durant SM (2015) Monitoring rarity: the critically endangered Saharan cheetah as a flagship species for a threatened ecosystem. PLoS One 10: e0115136.

85. Allen ML, Elbroch LM, Casady DS, Wittmer HU (2015) The feeding and spatial ecology of mountain lions (*Puma concolor*) in Mendocino National Forest, California. PLoS One 101: 51-65.

86. Bothma JDP, Le Riche EAN (1995) Evidence of the use of rubbing, scent-marking and scratching-posts by Kalahari leopards. Journal of Arid Environments 29: 511-517.

87. Gibson Hill CA (1950) Notes on the clouded leopard [*Neofelis nebulosa* (Griffith)]. Biodiversity Heritage Library 49: 543-546.

88. Mohamad SW, Rayan DM. Christopher WCT, Hamirul M, Mohamed A, et al. (2015) The first description of population density and habitat use of the mainland clouded leopard *Neofelis nebulosa* within a logged-primary forest in South East Asia. Population Ecology 57: 495-503.

89. Moo SSB, Froese GZ, Gray TN (2017) First structured camera-trap surveys in Karen State, Myanmar, reveal high diversity of globally threatened mammals. Oryx 51: 1-7.

90. O Brien SJ, Janice SM, Craig P, Lawrence H, Valerius V, Paul, et al. (1987) Biochemical genetic variation in geographic isolates of African and Asiatic lions. National Geographic Research 3: 114-124.

91. O Brien SJ, Johnson WE (2005) Big cat genomics. Annual Review 6: 407-429.

92. Tempa T, Hebblewhite M, Mills LS, Wangchuk TR, Norbu N, Wangchuk et al. (2013) Royal Manas National Park, Bhutan: a hot spot for wild felids. Oryx 47(2): 207-210.